\documentclass[a4paper]{PoS}

\usepackage{textcomp}
\usepackage{amsfonts} 
\usepackage{amssymb} 
\usepackage{amsmath} 
\usepackage{slashed}
\usepackage{graphicx}
\usepackage{pifont}
\usepackage{multirow,lscape}
\usepackage{array}
\usepackage{longtable}
\usepackage[FIGBOTCAP,TABBOTCAP]{subfigure}
\usepackage{verbatim}
\usepackage{bm}
\usepackage{comment}
\usepackage{xcolor}
\usepackage{url} 

\graphicspath{{./figs/}}

\newcolumntype{C}[1]{>{\centering\arraybackslash}p{#1}}

\title{ Performance of GTX Titan X GPUs and Code Optimization}

\ShortTitle{Performance of GTX Titan X GPUs}

\author{%
  Hwancheol Jeong, Sangbaek Lee, Weonjong Lee, \speaker{Jeonghwan Pak} \\
  Lattice Gauge Theory Research Center, CTP, and FPRD, \\
  Department of Physics and Astronomy, \\  
  Seoul National University,
  Seoul, 151-747, South Korea\\
  E-mail: \email{wlee@snu.ac.kr}
}

\author{%
  Jangho Kim \\
  National Institute of Supercomputing and Networking,
  Korea Institute of Science and Technology Information,
  Daejeon, 34141, South Korea \\
}

\author{%
  Juhyun Chung \\
  Hankuk academy of foreign studies, 
  Yongin, 449-854, South Korea \\
}

\abstract{ Recently Nvidia has released a new GPU model: GTX Titan X
  (TX) in a linage of the Maxwell architecture.  We use our conjugate
  gradient code and non-perturbative renormalization code to measure
  the performance of TX. The results are compared with those of GTX
  Titan Black (TB) in a lineage of the Kepler architecture.  We
  observe a significant gain in the single and double precision
  calculations much greater than the theoretical expectation.  }

\FullConference{
  The 33rd International Symposium on Lattice Field Theory\\
  14-18 July, 2015
  Kobe International Conference Center}

\begin{document}

\section{Introduction}
In 2015, NVIDIA released a new GPU of the Maxwell architecture: GTX
Titan X (TX).
TX GPUs have 20\% higher computing power in the single precision (SP)
floating point calculation than GTX Titan Black (TB) GPUs of the
Kepler architecture.
However, the former GPUs have significantly less computing power in
the double precision (DP) floating point calculation than the latter
GPUs by a factor of 1/7, if we compare their peak performance.
Here, we would like to compare the actual performance of TX GPUs with
that of TB GPUs.
For this purpose, we use our conjugate gradient (CG) inverter code to
check the SP performance of TX GPUs, and our non-perturbative
renormalization (NPR) code to probe the DP performance.
The CG code adopts the mixed precision algorithm. Hence, the SP
calculation is dominant in this code.
The NPR code calculates the matching factors in DP without much
network communication.
Hence, this code is good for testing the DP performance.
\section{GTX Titan X}

\begin{table}[h!]
\centering
%
\resizebox{1.0\textwidth}{!}{
  \begin{tabular}{|>{\centering\arraybackslash\small}m{3cm}
    ||>{\centering\arraybackslash\small}m{2.3cm}
    |>{\centering\arraybackslash\small}m{2.3cm}
    |>{\centering\arraybackslash\small}m{2.3cm}
    |>{\centering\arraybackslash\small}m{2.3cm}|}
  \hline
    \hline
    Architecture  & Fermi                      & \multicolumn{2}{c|} {Kepler} & Maxwell \\
    \hline
    GPU Model    & GTX 580                    & TITAN BLACK      & K40       & TITAN X \\
    \hline
    \hline
    SP TFLOPs & 1.58           & \color{blue}5.00 & 4.29      & \color{blue}6.00 \\
    \hline
    DP TFLOPs & 0.20           & \color{blue}1.30 & 1.43      & \color{blue}0.19 \\
    \hline
    Memory Size {\scriptsize (GB)}        & 1.5 &  6  & 12  & 12  \\
    \hline
    L1 cache$^1$
    \scriptsize (KB) & 64 & 64 & 64 & 96 \\
    \hline
    L2 cache \scriptsize (KB) & 768 & 1536 & 1536 & 3072 \\
    \hline
    Memory Bandwidth \scriptsize (GB/sec) & 192.4 & 336 & 288 & 336.5 \\
    \hline
    \hline
    \end{tabular}
}
\caption{Chip specification of recent NVIDIA GPUs. For more details,
  please refer to Refs.~\cite{ nvidia:titanx, nvidia:titanb,
    Jang:2014mxa}. }
\label{tab:spec}

\end{table}

In Table \ref{tab:spec}, we list chip specification for recent NVIDIA
GPU models which we have been using for our numerical study.
In the table, there are three generations of GPU chip architecture:
Fermi (2010), Kepler (2012), and Maxwell (2014).
TX GPUs are designed based on the Maxwell architecture, which have
twice more memory, twice more L2 cache and 50\% more L1
cache\footnote{ Here, the nominal L1 cache means the memory shared by
  the L1 cache and the shared memory.} than TB GPUs, while the memory
bandwidth is about the same.
One merit for TX is that its SP peak performance is about 20\% higher
than that of TB.
One superficial caveat is that the peak speed of the DP calculation in
TX GPUs is 1/7 of that of TB GPUs.
We will address this issue later in Section \ref{sec:npr}.

\section{CG Performance}
\label{sec:cg}

The conjugate gradient (CG) code uses the mixed precision algorithm
for HYP-smeared staggered quarks.
Hence, the SP calculation in this code is absolutely dominant compared
with the DP part.
This code performs about 15000 iterations in our production job for
$B_K$ on the $20^3\times 64$ MILC asqtad coarse lattices.
For each iteration, it uses network communication through infiniband
to transfer the data vector to the nearest neighbor nodes with 2 GPUs
per node.
\begin{figure}[t!]
  \centering
  \vspace{-7mm}
  \includegraphics[width=0.6\textwidth]{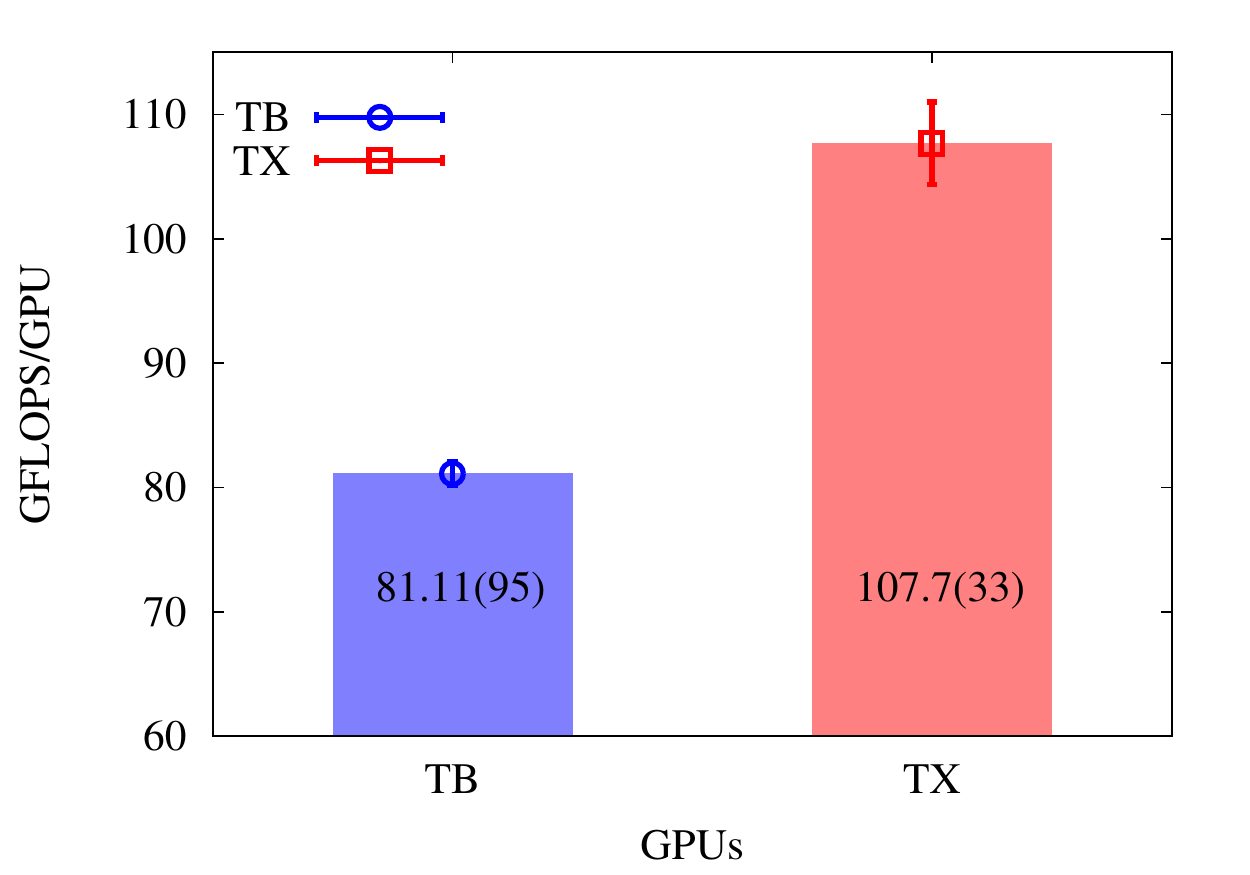}
  \caption{ CG performance on a single GPU in units of GFLOPS. The
    results are measured by running the CG inverter code 10 times on
    the $ 20^3 \times 64 $ MILC asqtad lattice. }
  \label{fig:cgperf}
\end{figure}

In Fig.~\ref{fig:cgperf}, we present the SP performance of the CG code
on a single TB or TX GPU obtained using the same production job for
$B_K$.
Here, since we use only a single GPU for this test job, there is no
network communication involved through the infiniband.
Theoretically, the ratio of the SP peak performance between TB and TX
GPUs is 1.2 as one can see in Table \ref{tab:spec}.
Hence, we expect that the SP performance will increase by 20\% for TX.
In practice, we obtain about 33(4)\% gain in the SP performance, which
is significantly higher than the theoretical expectation.
%

\begin{figure}[t!]
  \centering
  \vspace{-7mm}
  \includegraphics[width=0.7\textwidth]{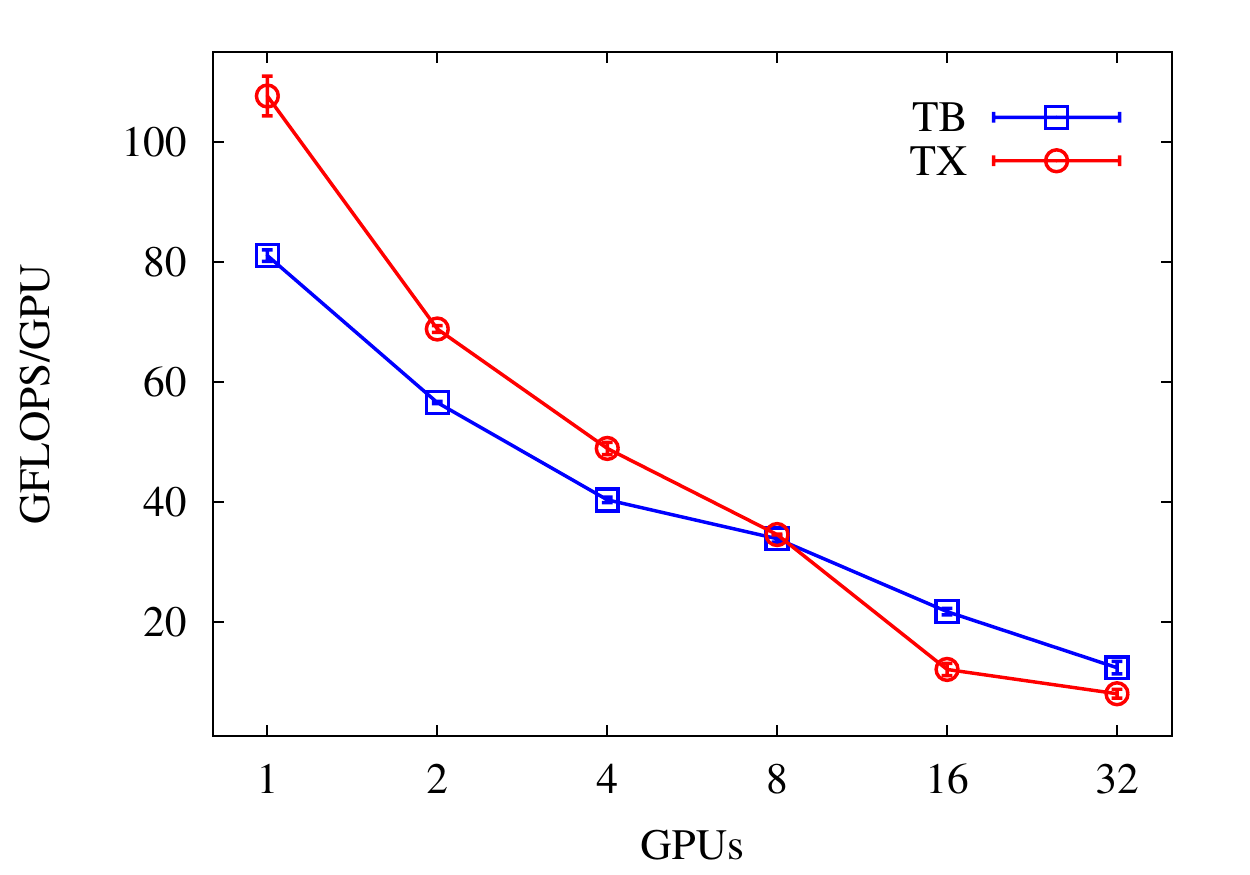}
  \caption{ Scalability of CG performances in the unit of GFLOPS with
    multiple GPUs. The results are obtained by running the CG code 10
    times on the $20^3 \times64$ MILC asqtad lattice.  }
  \label{fig:cg_sc}
\end{figure}

In Fig.~\ref{fig:cg_sc}, we repeat the same kind of performance tests
using the same production job on multiple GPUs of TB or TX type.
Here, we observe TX GPUs perform significantly better than TB GPUs
when we use the small number of GPUs ($\le 8$).
However, for more than 8 GPUs, TB GPUs outperform TX GPUs.
This indicates that we need to feed enough SP calculations above a
certain threshold (= 20 million SP floating point calculations per GPU
per iteration in CG) in order to obtain a significant gain using TX
GPUs compared with TB GPUs.
In Table \ref{tab:cg_sc}, we present numerical values of the data
points in Fig.~\ref{fig:cg_sc}.
\begin{table}[h!]
  \centering
  \resizebox{1.0\textwidth}{!}{
    \begin{tabular}{|>{\centering\arraybackslash\small}m{4cm}
      ||>{\centering\arraybackslash\small}m{2.5cm}
      |>{\centering\arraybackslash\small}m{2.5cm}
      ||>{\centering\arraybackslash\small}m{2.5cm}|}
    \hline
    \hline
    \# of nodes   & TB & TX   &  TX/TB ratio \\
    \hline
    \hline
    1    & 81.11(95) & 107.7(33)  & 1.328(44) \\
    \hline
    2    & 55.86(57) & 68.86(58)  & 1.233(16) \\
    \hline
    4    & 40.38(44) & 48.9(10)   & 1.212(29) \\
    \hline
    8    & 33.92(55) & 34.614(68) & 1.020(17) \\
    \hline
    16   & 21.74(55) & 12.1(10)   & 0.566(50) \\
    \hline
    32   & 12.4(11)  & 8.07(73)   & 0.650(81) \\
    \hline
    \hline
    \end{tabular}
  }
  \caption{
Data for CG performance on multiple GPUs.
They are measured by running the CG code 10 times on $ 20^3 \times 64
$ MILC asqtad lattice.
  }
  \label{tab:cg_sc}
\end{table}

Therefore, in order to maximize the computing performance per GPU, it
is better to use smaller number of GPUs for a given production job.
However, there are some caveats in this business: first, the memory of
TX GPUs are limited, and second, we need to make the running time of
the production job less than 2 hours in order to avoid hardware
failure as much as possible.
Hence, it is necessary to run the production job on multiple GPUs
which are usually greater than the smallest number of GPUs.
For example, when we run the production job on the $28^3 \times 96$
MILC asqtad fine lattices, it is ideal to use 4 TX GPUs in practice.

One ambiguity in this test is that the CPU configuration for TX GPUs
is different from that for TB GPUs.
For TX GPUs, we use i7-5930K CPUs with 32 giga byte DDR4 RAM.
For TB GPUs, we use i7-4820K CPUs with 32 giga byte DDR3 RAM.
DDR4 is slightly faster than DDR3.
In addition, i7-5930K CPU is slightly faster than i7-4820K CPU.
Hence, this could make a tiny difference in the above data shown in
Figs.~\ref{fig:cgperf} and \ref{fig:cg_sc}.

\section{NPR performance}
\label{sec:npr}

Here we present the performance of the non-perturbative
renormalization (NPR) code in the RI-MOM scheme.
The NPR code use only the DP floating point calculation and does not
have any network communication with nearest neighbors during the NPR
calculation.
Hence, the NPR code is very adequate to measure the DP performance of
GPUs.

In the NPR code, the subroutine to calculate the one color trace
four-fermion operator contraction occupies 97\% of the total running
time of the NPR code \cite{Jeong:2014jia}.
\begin{equation}
  \label{eq:ffop}
  \begin{split}
    O^{f_1 f_2 f_3 f_4}_{i;I} (z) &=
    [\overline{\chi}^{f_1}_{i;{\color{red}c_1}}(z_A) \overline{(\gamma_{S_1} 
      \otimes \xi_{F_1})}_{AB} \chi^{f_2}_{i;{\color{red}c_2}}(z_B) ]
    [\overline{\chi}^{f_3}_{i;{\color{blue}c_3}}(z_C)
    \overline{(\gamma_{S_2} \otimes \xi_{F_2})}_{CD}
    \chi^{f_4}_{i;{\color{blue}c_4}}(z_D)] \\
    & \hspace{5pc} \times [U_{i;AD}]_{{\color{red}c_1}
      {\color{blue}c_4}} (z) [U_{i;CB}]_{{\color{blue}c_3} {\color{red}c_2}}
    (z) \\
  \end{split}
\end{equation}
We present a general form of one-color trace four-fermion operators
in Eq.~\eqref{eq:ffop}.
Here, $\chi$ and $\bar\chi$ represent staggered quark fields, and
$U_{i;AD}(z)$ represents a gauge link field.
The subscripts $S_1$ and $S_2$ represent the spin structure and
the subscripts $F_1$ and $F_2$ represent the taste structure.
In order to measure the DP performance of GPUs, we use this
subroutine which calculate the one color trace four-fermion
contraction.

\begin{figure}[t!]
  \centering
  \vspace{-7mm}
  \includegraphics[width=0.6\textwidth]{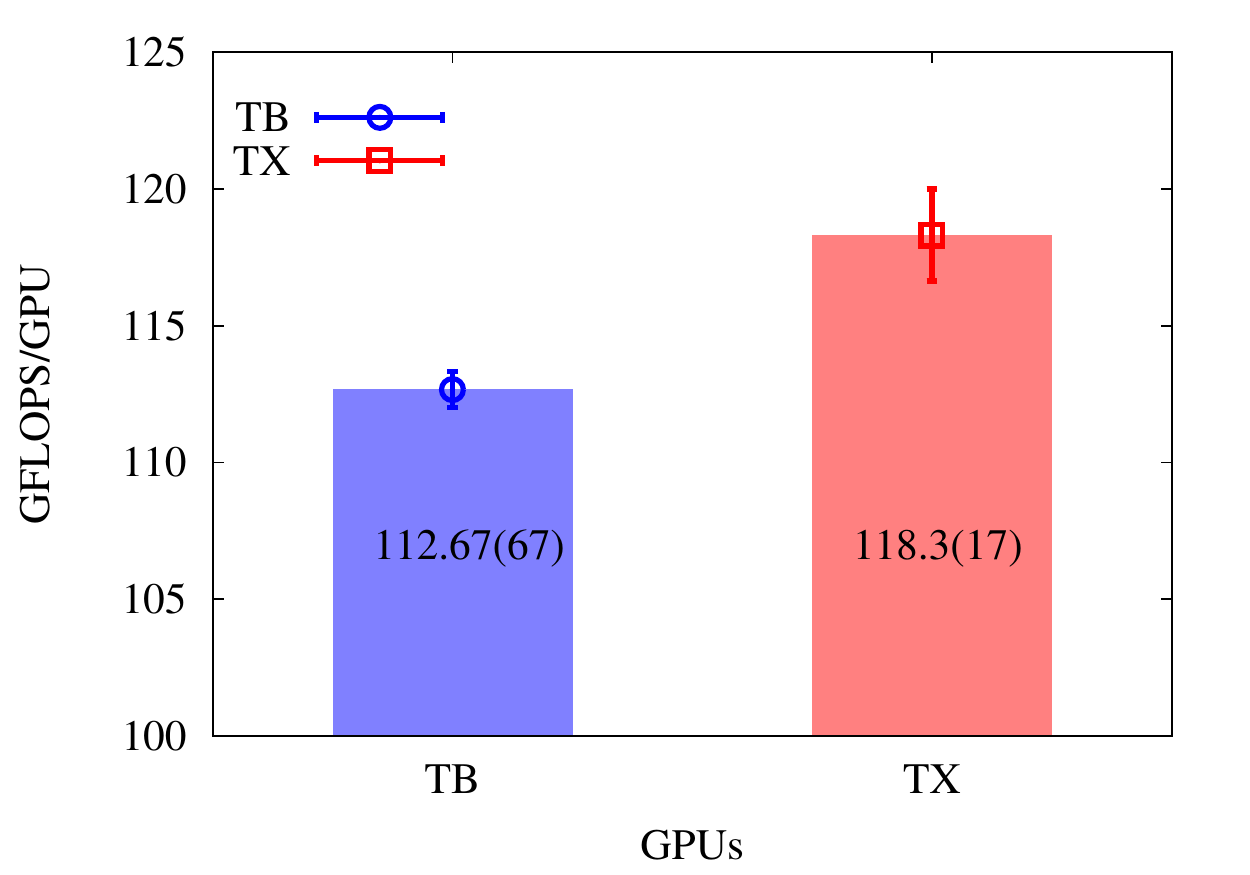}
  \caption{ NPR performance, actually DP performance, in the unit of
    GFLOPS. The results are measured by running the NPR code 10 times
    on the $ 20^3 \times 64 $ MILC asqtad lattice at $ a \approx
    0.12fm $.  }
  \label{fig:npr_perf}
\end{figure}
As one can see in Table \ref{tab:spec}, the peak DP performance for TX
is 1/7 of that for TB theoretically.
Hence, theoretically we expect that the NPR code runs 7 times faster
on a TB GPU.
In Fig.~\ref{fig:npr_perf}, we present the DP performance of the NPR
code on TB and TX GPUs.
To our big surprise, it turns out that the TX GPU outperform the TB
GPU by 5.0(6)\%.

What is the reason for this unexpected surprise?
The answer is that our NPR code is dominated by the data transfer
between the GPU global memory and the CUDA cores.
Hence, even though the TB GPU has 7 times more DP calculation power
than TX, the TB GPU cannot feed the data into the CUDA cores as fast.
The bottle neck is on the data transfer in the case of the NPR code.

Let us address this issue on the bottle neck in data transfer more
systematically.
Basically we want to introduce a concept of the CGMA ratio (Compute to
Global Memory Access) \cite{Jeong:2013vqa}.
The CGMA ratio is the number of DP floating point calculations per
memory access.
For example, let us consider the calculation of $C = A+B$.
There are three memory accesses to read data from $A$ and $B$ and
to save data into $C$.
There is one DP floating point calculation for $+$.
Hence, the CGMA ratio for this calculation is $1/3$.

\begin{table}[h!]
  \centering
  %
  \resizebox{1.0\textwidth}{!}{
    \begin{tabular}{|>{\centering\arraybackslash\small}m{2cm}
        ||>{\centering\arraybackslash\small}m{3cm}
        |>{\centering\arraybackslash\small}m{3cm}
        |>{\centering\arraybackslash\small}m{4cm}
        |>{\centering\arraybackslash\small}m{2.5cm}|}
        \hline
        \hline
        GPU Model & Memory bandwidth (GB/sec) & Peak performance (GFLOPS)
        & Memory bandwidth (GB/sec) for CGMA=1 & CGMA for peak performance \\
        \hline
        \hline
        GTX 580     & 192.4 &   204.8 &  1638.4 &  8.5 \\
        \hline
        TB          & 336   &  1331.2 & 10649.6 & 31.7 \\
        \hline
        K40         & 288   &  1464.3 & 11714.6 & 40.7 \\
        \hline
        TX          & 336.5 &   194.6 &  1556.5 &  4.6 \\
        \hline
        \hline
      \end{tabular}
  }
  \caption{ CGMA Ratio for various GPUs in units of
    [GFLOPS/(GB/sec)*Byte] }
  \label{tab:cgma}
\end{table}

In Table \ref{tab:cgma}, we present data for the CGMA ratio on various
GPUs.
Here, we present the memory bandwidth and the peak DP performance
in the second and third columns, respectively.
In the fourth column, we present memory bandwidth required to
achieve the CGMA ratio equal to 1.
In other words, with this bandwidth we can achieve one to one ratio
between the DP floating point calculation and the data transfer.
In the fifth column, we show the CGMA ratio for the peak performance
on each GPU.

The CGMA ratio of our NPR code is 2.96 [GFLOPS/(GB/sec)*Byte].
Let us assume that we use the full memory bandwidth of each GPU.
Then, we can achieve 124 GFLOPS for TB and 125 GFLOPS for TX at CGMA =
2.96.
Hence, we observe that this actual DP performance is much lower than
the theoretical peak.
This indicates that the bottle neck is on the data transfer.

Then at CGMA = 2.96, we expect that TX GPUs will outperform TB GPUs
only by 0.15\%.
However, in practice we find a gain of 5.0(6)\% in
Fig.~\ref{fig:npr_perf}, which is much larger than the theoretical
expectation by an order of magnitude.
At present, we do not understand this enormous amount of gain in TX.
This needs further investigation in the future.

\begin{figure}[t!]
  \centering
  \vspace{-7mm}
  \includegraphics[width=0.7\textwidth]{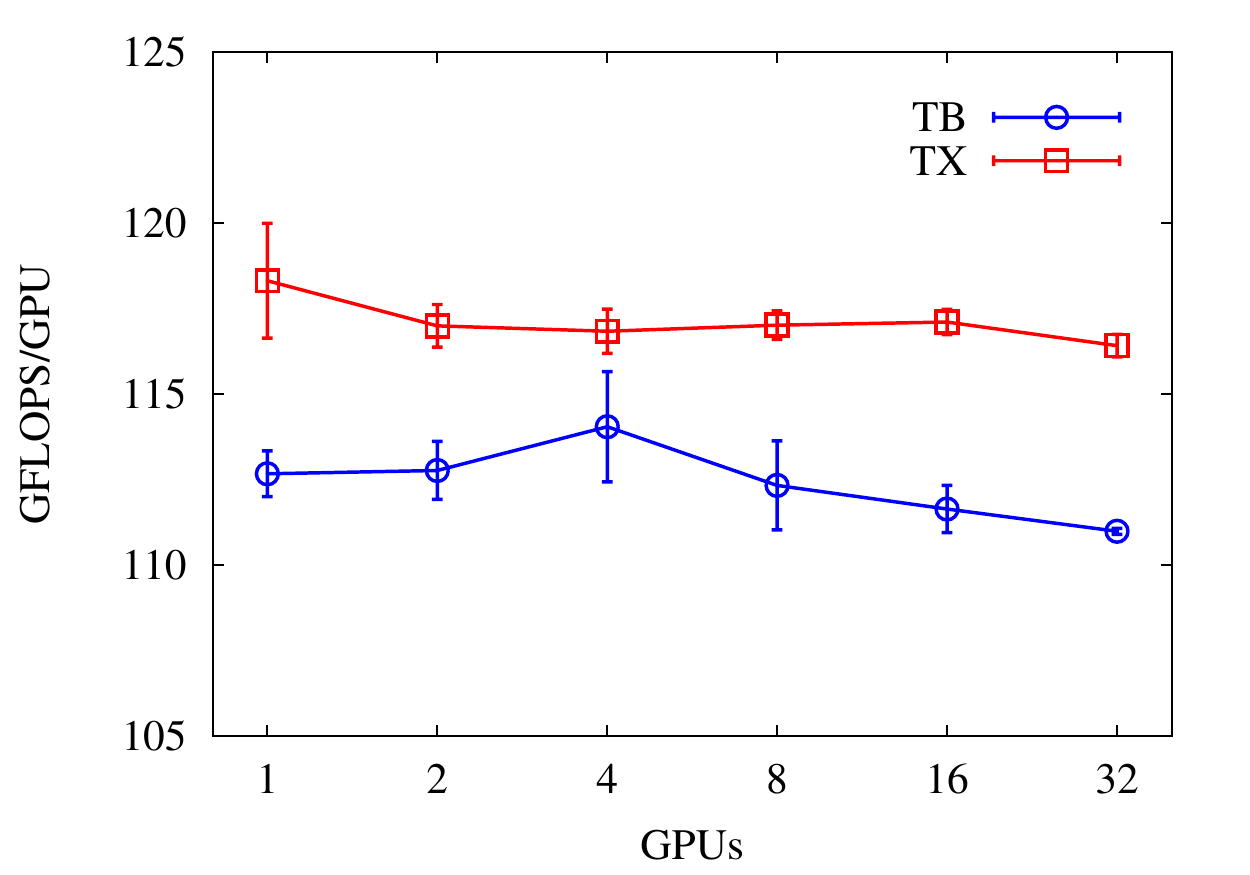}
  \caption{ Scalability of the DP performance on multiple GPUs.  We
    use the NPR code to measure the DP performance on TX or TB GPUs.
    on the $20^3\times 64$ MILC asqtad lattices. }
  \label{fig:npr_sc}
\end{figure}

In Fig.~\ref{fig:npr_sc}, we present the DP performance per GPU
when we run the NPR code on multiple GPUs.
In Table \ref{tab:npr_sc}, we summarize the actual values of the
data points in Fig.~\ref{fig:npr_sc}.
Since the NPR code does not have data transfer to the nearest
neighbors, we expect that the performance will be constant
regardless of the number of GPUs used for the measurement.
The data in Fig.~\ref{fig:npr_sc} is consistent with this expectation.
In this test jobs, TX GPUs outperform TB GPUs by about 5\% regardless
of the number of GPUs.
This large gain remains as a conundrum, which needs further
investigation in the future.

In our production job on NPR, we use 28 GPUs of TX or TB type on
$28^3\times 96$ MILC asqtad lattices.
A single production job takes about 6.5 hours.
\begin{table}[h!]
  \centering
  %
  \resizebox{1.0\textwidth}{!}{
    \begin{tabular}{|>{\centering\arraybackslash\small}m{4cm}
          ||>{\centering\arraybackslash\small}m{2.5cm}
          |>{\centering\arraybackslash\small}m{2.5cm}
          ||>{\centering\arraybackslash\small}m{2.5cm}|}
        \hline
        \hline
        \# of nodes      &        TB   & TX         & TX/TB ratio \\
        \hline
        \hline
        1                & 112.67(67)  & 118.3(17)  & 1.050(6)  \\
        \hline
        2                & 112.23(79)  & 117.04(57) & 1.038(9)  \\
        \hline
        4                & 114.1(16)   & 116.84(65) & 1.024(6)  \\
        \hline
        8                & 112.3(13)   & 117.02(42) & 1.042(4)  \\
        \hline
        16               & 111.64(69)  & 117.11(37) & 1.049(7)  \\
        \hline
        32               & 110.99(90)   & 116.42(33) & 1.049(3)  \\
        \hline
        \hline
    \end{tabular}
  }
  \caption{ DP performance per GPU in units of GFLOPS. The data
  describes the data points in Fig.~\protect\ref{fig:npr_sc}.}
  \label{tab:npr_sc}
\end{table}

\section{Conclusion}

Recently NVIDIA has released a new GPU model, TX.
Here, we present the results of our performance test on TX GPUs and
compare them with those on TB GPUs.

In the SP performance test, TX outperforms TB by 33(4)\%, which is
significantly higher than the theoretical expectation.
This test is done using the CG code.

In the DP performance test with the NPR code, TX outperforms TB by
5.0(6)\%, which is dramatically larger than the theoretical
expectation of 0.15\% gain.
At present, we do not understand this gain very well.
This needs further investigation in the future.
The DP performance of TX with the NPR code is controlled by
the data transfer, which is the main bottle neck.
Hence, in order to increase the DP performance, we need to
optimize the NPR code further for higher CGMA ratio.

\acknowledgments
The research of W.~Lee is supported by the Creative Research
Initiatives program (No.~2015001776) of the NRF grant funded by the
Korean government (MSIP).
This work was supported by SNU Undergraduate Research Program. W.~Lee
would like to acknowledge support from the KISTI supercomputing center
through the strategic support program (No.KSC-2014-G3-002) with much
gratitude.
J.~Kim was supported by a Young Scientists Fellowship through the
National Research Council of Science \& Technology (NST) of Korea.
Computations were carried out on the DAVID GPU clusters at Seoul
National University.
%


\bibliography{refs}

\providecommand{\href}[2]{#2}\begingroup\raggedright\begin{thebibliography}{1}

\bibitem{nvidia:titanx}
{\em {GeForce GTX TITANN X Specifications}}.
\newblock
  \url{http://www.geforce.com/hardware/desktop-gpus/geforce-gtx-titan-x/specifications}.

\bibitem{nvidia:titanb}
{\em {GeForce GTX TITANN X Specifications}}.
\newblock
  \url{http://www.geforce.com/hardware/desktop-gpus/geforce-gtx-titan-black/specifications}.

\bibitem{Jang:2014mxa}
{\bf SWME} Collaboration, Y.-C. Jang, H.~Jeong, J.~Kim, W.~Lee, J.~Pak, and
  Y.~Chung, {\it {Code Optimization on Kepler GPUs and Xeon Phi}},  {\em PoS}
  {\bf LATTICE2014} (2014) 035, [\href{http://xxx.lanl.gov/abs/1411.2223}{{\tt
  1411.2223}}].

\bibitem{Jeong:2014jia}
{\bf SWME} Collaboration, H.~Jeong, J.~Kim, J.~Kim, W.~Lee, J.~Pak, and
  S.~Park, {\it {Non-perturbative Renormalization of Four-Fermion Operators
  Relevant to $B_K$ with Staggered Quarks}},  {\em PoS} {\bf LATTICE2014}
  (2015) 286, [\href{http://xxx.lanl.gov/abs/1410.6607}{{\tt 1410.6607}}].

\bibitem{Jeong:2013vqa}
H.~Jeong, W.~Lee, J.~Pak, K.-j. Choi, S.-H. Park, J.-s. Yoo, J.~H. Kim, J.~Lee,
  and Y.~W. Lee, {\it {Performance of Kepler GTX Titan GPUs and Xeon Phi
  System}},  {\em PoS} {\bf LATTICE2013} (2014) 423,
  [\href{http://xxx.lanl.gov/abs/1311.0590}{{\tt 1311.0590}}].

\end{thebibliography}\endgroup

\end{document}